\documentclass{article}
\usepackage[T1]{fontenc} % add special characters (e.g., umlaute)
\usepackage[utf8]{inputenc} % set utf-8 as default input encoding
\usepackage{ismir,amsmath,cite,url}
\usepackage{graphicx}
\usepackage{color}

\usepackage{subcaption}
\usepackage[ruled,vlined,linesnumbered]{algorithm2e}
\usepackage{booktabs}
\usepackage{multirow}

\usepackage{amssymb}
\usepackage{bbm}

\usepackage{lineno}
\title{AccoMontage2: A Complete Harmonization and Accompaniment Arrangement System}

\multauthor
{Li Yi$^{1, 2}$ \hspace{1cm} Haochen Hu$^{1, 2}$ \hspace{1cm} Jingwei Zhao$^{3}$ \hspace{1cm} Gus Xia$^{1, 2}$} {
$^1$ Music X Lab, NYU Shanghai  \hspace{1cm} $^2$ MBZUAI\\
$^3$ Institute of Data Science, NUS\\
{\tt\small 
\href{mailto:ly1387@nyu.edu}{ly1387@nyu.edu}, \href{mailto:hh1933@nyu.edu}{hh1933@nyu.edu}, \href{mailto:jzhao@u.nus.edu}{jzhao@u.nus.edu}, \href{mailto:gxia@nyu.edu}{gxia@nyu.edu}}
}

\def\authorname{L. Yi, H. Hu, J. Zhao, and G. Xia}

\usepackage[bookmarks=false,pdfauthor={\authorname},pdfsubject={\papersubject},hidelinks]{hyperref}
\begin{document}

\maketitle
\begin{abstract}
We propose AccoMontage2, a system capable of doing full-length song harmonization and accompaniment arrangement based on a lead melody.\footnote{Codes and dataset at \href{https://github.com/billyblu2000/accomontage2}{https://github.com/billyblu2000/accomontage2}.} Following AccoMontage, this study focuses on generating piano arrangements for popular/folk songs and it carries on the generalized template-based retrieval method. The novelties of this study are twofold. First, we invent a harmonization module (which AccoMontage does not have). This module generates structured and coherent full-length chord progression by optimizing and balancing three loss terms: a micro-level loss for note-wise dissonance, a meso-level loss for phrase-template matching, and a macro-level loss for full piece coherency. Second, we develop a graphical user interface which allows users to select different styles of chord progression and piano texture. Currently, chord progression styles include Pop, R\&B, and Dark, while piano texture styles include several levels of voicing density and rhythmic complexity. Experimental results show that both our harmonization and arrangement results significantly outperform the baselines. Lastly, we release AccoMontage2 as an online application as well as the organized chord progression templates as a public dataset. 
\end{abstract}

\section{Introduction}\label{sec:introduction}
Accompaniment arrangement is a difficult music generation task involving structured constraints of melody, harmony, and accompaniment texture. A high-quality arrangement could help with various downstream tasks and applications, such as compositional style transfer \cite{wang2020learning}, automatic accompaniment \cite{xia_2018}, and score-informed source separation and music synthesis \cite{DBLP:conf/ismir/LinXKJ21}. 

As one of the most promising arrangement systems, AccoMontage \cite{zhao2021accomontage} uses a generalized template-based approach to first search for roughly-matched accompaniment phrases as the reference and then re-harmonize the selected reference via style transfer. It generates much more coherent results than purely learning-based algorithms, especially for full-length song arrangements. % to better match the query lead sheet

However, AccoMontage is not yet a ``complete'' accompaniment generation system in the strict sense, as it still calls for chord input from users and cannot harmonize a melody. To this end, we develop AccoMontage2, a system capable of full-length song harmonization and accompaniment arrangement based on a lead melody by equipping AccoMontage with two extra components: 1) a novel harmonization module, and 2) a graphical user interface.

The main novelty of our system lies in the harmonization module. We first collect a high-quality chord progression dataset and re-organize the phrases with respect to different styles to serve as reference templates. Then, we use dynamic programming (DP) to generate structured and coherent chord progressions given a query lead melody with phrase annotation. Specifically, the DP algorithm optimizes a multi-level loss function consisting of three terms:  1) a micro-level loss for note-wise melody-chord matching, 2) a meso-level loss for phrase-template matching, and 3) a macro-level loss for the whole-piece coherency. The first term evaluates the dissonance between the melody and the candidate chords. The second term prefers chord progressions with the same length as the target melody phrases. The third term computes how well the candidate phrases connect with each other to form an organic whole. Experimental results show that both our harmonization and arrangement results significantly outperform the baselines. 

In addition, we develop a graphical user interface which allows the user to select different styles of chord progression and piano texture. Currently, chord progression styles include R\&B, Dark, Pop-standard, and Pop-complex. Piano texture styles include several levels of voicing density and rhythmic complexity. 

We release the AccoMontage2 as an online application\footnote{Online GUI link at \href{https://billyyi.top/accomontage2}{https://billyyi.top/accomontage2}.} as well as the organized chord progression templates as an open-source dataset. 

In brief, the contributions of our paper are as follows:
\begin{itemize}
    \item A complete system for full-length song harmonization and accompaniment arrangement;% based on a lead melody;
    \item An effective harmonization algorithm with state-of-the-art  performance;
    \item A graphical user interface for controllable piano accompaniment generation.
\end{itemize}
%(1) A full-length song harmonization and accompaniment arrangement based on a lead
%(2) An effective state-of-the-art harmonization algorithm.
%(3) A graphical user interface for controllable arrangement.

%
\section{Related Works}
%We review two topics related to our work: melody harmonization and accompaniment arrangement.

\subsection{Melody Harmonization}
Melody harmonization refers to the task of generating a harmonic chordal accompaniment for a given melody \cite{chuan2007hybrid, simon2008mysong}. It has been typically formulated as a prediction task, \textit{i.e.}, to predict a sequence of chord labels conditioned on the lead melody. Recent mainstream methods range from hidden Markov models \cite{tsushima2017function, tsushima2018generative} to deep neural networks \cite{lim2017chord, yeh2021automatic, sun2021melody}. Such models are typically trained to fit a groundtruth melody-chord mapping, but do not account for the fact that one melody can be harmonized with various styles in terms of genre, chord complexity, \textit{etc}. In fact, the current state-of-the-art models \cite{yeh2021automatic, sun2021melody} only support simple triads and up to a few common seventh chords. Also, predictions are made locally, where neither phrase-level progression nor inter-phrase structures are explicitly considered.

In this paper, we re-formulate melody harmonization with a novel template matching approach. The usage of existing templates for music generation has been a popular idea. Existing template-based methodologies include learning based unit selection \cite{bretan2016unit,xia_2018}, rule-based score matching  \cite{chen2013automatic, wu2016emotion}, and genetic algorithms \cite{liu2012polyphonic}. In our case, we match the lead melody with chord templates from a library based on rule-based criterion and subject to user control. Such an idea is inspired by the fact that music producers tend to pick up off-the-shelf chord templates instead of harmonizing from scratch. In addition, they also have control on what style of the chords to use.

Existing template-matching attempts for harmonization typically focus on half-bar level \cite{fujishima1999real}. In contrast,  our model deals with phrase-level matching. We design three loss terms that measure melody-chord fitness at note-wise, intra-phrase, and inter-phrase levels, respectively. Our chord library is finely organized, supporting up to ninth chords with voice leading and various genres. Our model can therefore generate structured and coherent full-length chord progressions with different styles.

\subsection{Accompaniment Arrangement}
The task of accompaniment arrangement aims to generate an accompaniment conditioned on a given lead sheet (\textit{i.e.}, a lead melody with chord progression). The quality of arrangement is related to chordal harmony, texture richness, and long-term structure. For this task, existing learning-based models often do well in harmony and texture, but are less capable of long-term generation \cite{dong2018musegan, liu2018lead, jia2019impromptu, wang2020learning, ren2020popmag, zhu2018xiaoice}. Previous template-matching models can easily maintain long-term structures, but suffer from fixed elementary textures and often fail to generalize \cite{chen2013automatic, wu2016emotion, liu2012polyphonic}. 

To break such a dilemma, the AccoMontage system \cite{zhao2021accomontage} introduces a generalized template-matching methodology, where phrase-level accompaniment templates are first searched by rule-based criterion, and then re-harmonized via deep learning-based style transfer. The search stage and the style transfer stage each optimize high-level structure and local coherency, thus guaranteeing the arrangement of high-quality accompaniment.

In this paper, we integrate our harmonization module with AccoMontage and upgrade it to a complete accompaniment generation model. An input melody is first harmonized with stylistic chord progression and then arranged with piano textures. With an additional GUI design, our model offers a flexible degree of control, including harmony styles and texture complexity.

\section{Methodology}

\begin{figure}[htb]
 \centerline{
 \includegraphics[width=0.95\columnwidth]{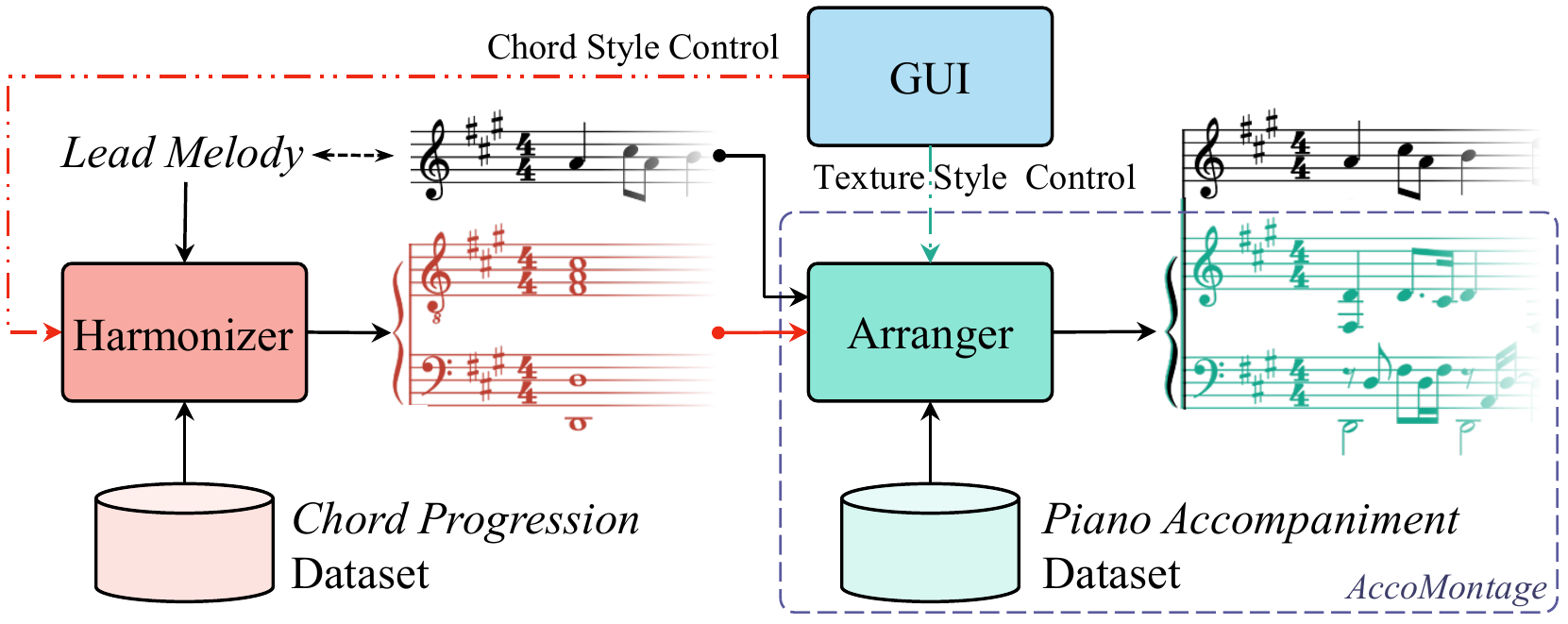}}
 \caption{Diagram of AccoMontage2 system.}
 \label{big_picture_diagram}
\end{figure}

The system diagram of our AccoMontage2 system is shown in Figure \ref{big_picture_diagram}. It can achieve full-length song harmonization and accompaniment arrangement. The input of the system is a query lead melody with phrase annotation. The harmonization module will first harmonize the melody. The generated chord progression will be sent into AccoMontage together with the original melody to arrange accompaniment. Lastly, a GUI is provided for users to adjust chord and accompaniment styles. 

For the rest of this section, we first introduce the structure of the dataset and how we re-organize it in Section 3.1. Then we describe the harmonization module in Section 3.2. After that, we provide an overview of the AccoMontage system in Section 3.3. Finally, we show how the GUI is constructed to enable style controllability in Section 3.4. 

\subsection{Dataset Curation}

A self-collected chord progression dataset is used as the reference templates for our harmonization algorithm. We create the dataset based on an existing chord progression collection \cite{kotoulas} that contains 64,524 MIDI files, most of which are chord progression tracks with different style labels. The original dataset \cite{kotoulas} cannot be adapted to our model directly for several reasons. First, some tracks are not pure chord progression, containing mixed melody segments. Second, many style labels are unnecessary. For example, two different styles may have similar musical elements that are hard to differentiate. Third, there are redundant progressions only different in their keys.

To solve the problems above, we process and re-organize the dataset as follows. First, we remove the MIDI files that contain melody segments and complex rhythmic textures. Second, based on our subjective assessment, we manually define a style mapping function to map the original style labels to newly defined ones.\footnote{A detailed description of style mapping is provided in our \href{https://github.com/billyblu2000/accomontage2}{dataset}.} The newly defined styles are: ``Pop-standard'', ``Pop-complex'', ``Dark'', and ``R\&B''. While Pop-standard contains only triad chords, Pop-complex contains seventh, ninth, and sometimes even more chromatic ones. Note that some templates do not have an original style label and are thus labeled as “Unknown”. Third, we remove all the redundant progressions.

\begin{figure}[htb]
 \centerline{
 \includegraphics[width=0.95\columnwidth]{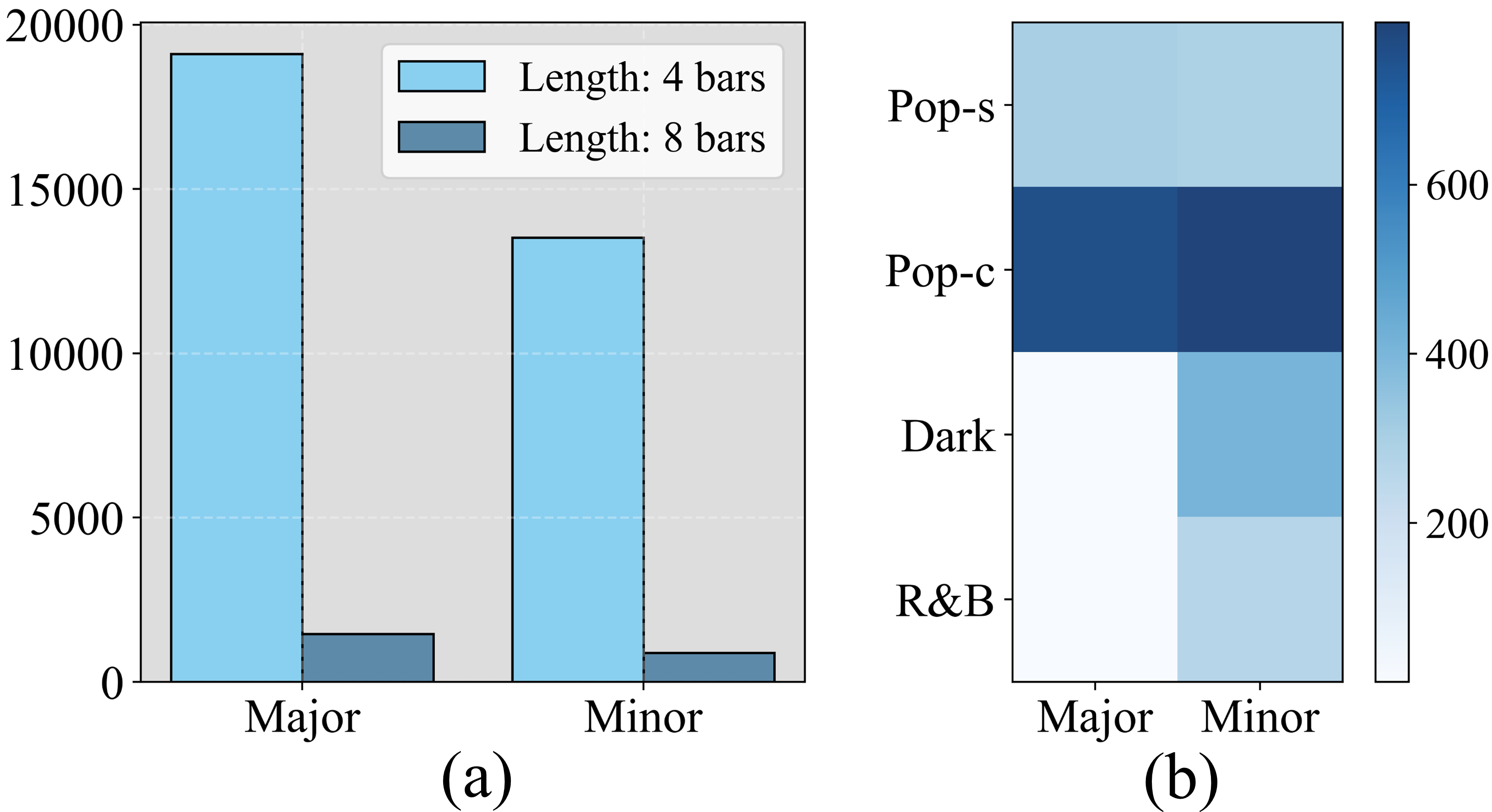}}
 \caption{Statistics of the curated dataset.}
 \label{dataset_statistics}
\end{figure}

The final curated dataset contains 5762 pieces of chord progression templates. Each template has 3 additional labels: 1) The length label. Templates are either of 4 bars or 8 bars in length in our dataset. Figure \ref{dataset_statistics} (a) shows a distribution of the length of the templates. 2) The mode label. We currently only label the mode of the templates as either major or minor. 3) The style label. As mentioned above, four styles in total are acquired. The distribution of different styles among modes are shown in Figure \ref{dataset_statistics} (b).

% \begin{table}[htbp]
%   \centering
%     \resizebox{.23\textwidth}{!}{
%     \begin{tabular}{lcccc}
%     \toprule
%           & \textbf{Mean} & \textbf{Variance} &  \\
%     \midrule
%     \textbf{MIDI Pitch}      & 57.00 & 167.70 \\
%     \textbf{MIDI Velocity}   & 79.05 & 457.89 \\
%     \textbf{Duration}   & 11.05 & 103.94  \\
%     \bottomrule
%     \end{tabular}%
%     }
%   \caption{Dataset Statistics of pitch, velocity and duration.}
%   \label{multi_level_score}%
% \end{table}%

Finally, we represent the reference template space as a collection of tuples:
\begin{equation}
    r = \{(r_m^{\text{chord}}, r_m^{\text{label}})\}_{m=1}^N,
\end{equation}
\noindent
where $r_m^{\text{chord}}$ and $r_m^{\text{label}}$ are the chord progression and the labels of the $i^{\text{th}}$ reference template; N is the volume of the reference space. $r_m^{\text{chord}}$ can be seen as a sequence of chords. We quantize and sample the chords at 8th notes from the original chord progression track. Each chord is represented as its root note and a label to indicate whether the corresponding triad is a major triad or a minor triad.

\subsection{Harmonization Module}\label{harmonization_module}
We design a harmonization model that generates structured and coherent full-length chord progression for a given lead melody with phrase annotation. The model takes a multi-phrase melody as input and outputs a list of optimal chord progression identities. Each identity contains a group of progressions that have the same Roman numeral sequence (\textit{e.g.}, I-vi-ii-V) but different styles (\textit{e.g.}, Pop-standard or R\&B) and are up for the user to choose.
The model considers three levels of losses that involve melody-chord correspondence and is optimized by a dynamic programming (DP) algorithm. Specifically, the DP algorithm optimizes a multi-level loss function consisting of three terms:  1) a micro-level loss for note-wise melody-chord matching, 2) a meso-level loss for phrase-template matching, and 3) a macro-level loss for the full piece coherency. 

\subsubsection{Micro-level loss
}
\begin{figure}[h]
\centering
\begin{subfigure}[b]{0.222\textwidth}
   \includegraphics[width=1\linewidth]{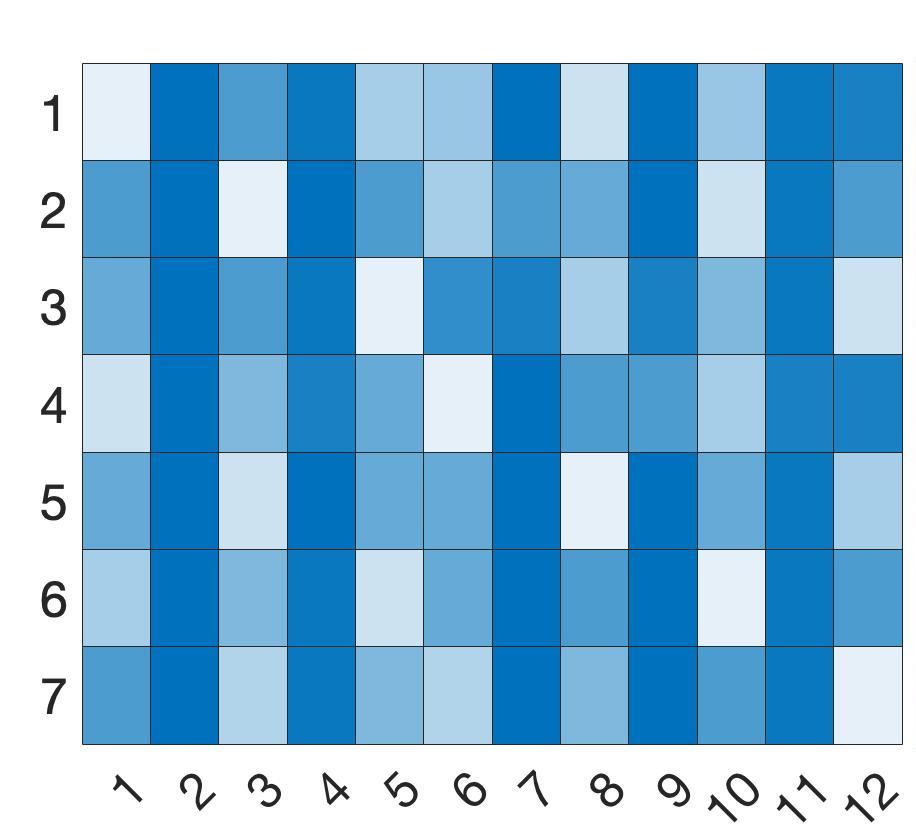}
   \caption{Major}
   \label{fig:Ng1} 
\end{subfigure}
\hfill
\begin{subfigure}[b]{0.249\textwidth}
   \includegraphics[width=1\linewidth]{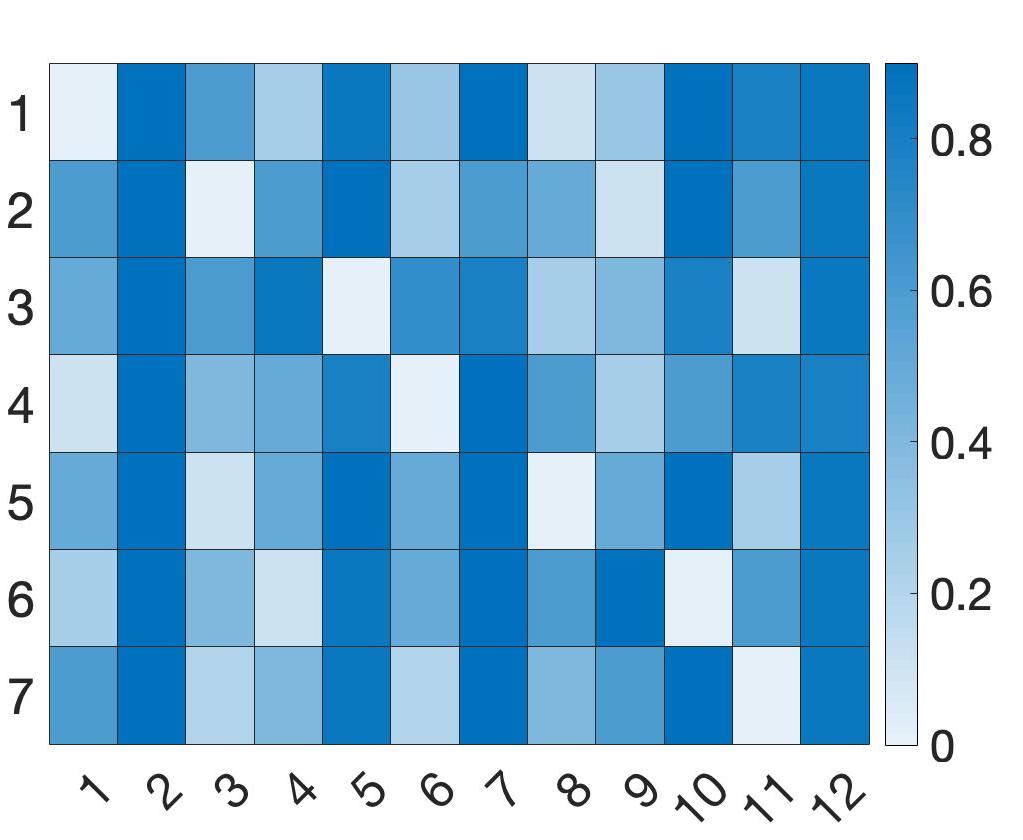}
   \caption{Minor}
   \label{fig:Ng2}
\end{subfigure}

\caption[]{Micro-level loss indicating note-wise melody-chord dissonance. }
\label{fig:micro-level-loss}
\end{figure}

The micro-level loss $L_{\text{mic}}$ computes the level of dissonance between a melody phrase and candidate progressions note by note. We mainly consider the interval between a melody note and the root of the chord in the corresponding position. The more dissonant the interval is (tritone, minor seconds, etc.), the higher the loss.  On the other hand, the more harmonious the interval is (unison, perfect fourth, perfect fifth, etc.), the lower the loss. In addition, the mode of the piece and the harmonic function of the chords will also affect the dissonance level. 
We refer to the ``rank order of consonances and their degree of recurrence'' \cite{minorloss} and design the micro-level loss shown in Figure \ref{fig:micro-level-loss}, where matrix (a) is for major mode and matrix (b) is for minor. For both matrices, each row represents the degree of the chord (we only consider diatonic chords, \textit{e.g.}, 1 for tonic and 5 for dominant) and each column represents the interval between the melody note and the key center. In Figure \ref{fig:micro-level-loss}, the darker the shade, the more dissonant we consider the pair of the melody note and chord is, in which case we set a higher micro-level loss.

For the $i^{\text{th}}$ phrase, $L_{\text{mic}}$ is computed by summing up the micro-level loss note by note and then dividing it by the length of the phrase. We further normalize $L_{\text{mic}}$ to the range of $[0, 1]$.

\subsubsection{Meso-level loss}\label{Meso-level-loss}
The meso-level loss $L_{\text{mes}}$ considers the integrity of a candidate. It encourages chord progressions with the same length as the target melody phrases and penalizes the rest. For a given phrase of melody, we consider chord progressions of the same length and we also concatenate shorter chord progressions to make them candidates. Here, we introduce a transition score to penalize the concatenated candidates. We concatenate two progressions $r_m^{\text{chord}}$ and $r_n^{\text{chord}}$ from the  reference template space, and the transition score considers the very two bars connecting them (\textit{i.e.}, the last bar of $r_m^{\text{chord}}$ and the first bar of $r_n^{\text{chord}}$). In specific, we count the occurrence of this two-bar chord progression in the dataset and then normalize it by applying a logarithmic function with $N$, the size of the reference template space as the base. Suppose the occurrence is $c\in\mathbb{N}$, the transition loss of this $m\rightarrow n$ concatenated candidate is defined as:
\begin{align}
T_{m\rightarrow n} := 1 + \log_{N}\frac{1}{c}
\end{align}

Moreover, as we do not wish to penalize the non-concatenated candidates, we introduce $\delta_1$ such that $\delta_1 = 1$ if the candidate is concatenated and $\delta_1 = 0$ otherwise. On the other hand, we do not wish to consider candidates with length not equal to the length of the phrase. Hence we introduce $\delta_2$ such that $\delta_2 = 1$ if the candidate has equal length as the melody phrase, and $\delta_2 = e^{-10}$ otherwise. The meso-level loss of the $i^{\text{th}}$ phrase is explicitly:
\begin{align}
L_{\text{mes}}^i := \delta_1 T_{m\rightarrow n} + (\frac{1}{\delta_2} - 1)
\end{align}

% Moreover, we search each candidate progression through the POP909 data set to find the best-matched phrase and its relative location in the corresponding piece. We penalize the candidate if it is not in the right location and call it the position loss. Finally, 

% \[
% L_{mes} = \alpha T_{i\rightarrow j} + (1-\alpha) \mbox{pos}
% \]
% and $T_{i\rightarrow j}=0$ if the candidate is not concatenated.

\subsubsection{Macro-level loss}

The macro-level loss $L_{\text{mac}}$ computes how well the candidate phrases connect with each other. We consider how smooth the transition is from the previous progression to the current progression using the same transition loss as in Section 3.2.2. For the $i^\text{th}$ phrase and its corresponding progression candidate, $L_{mac}^i$ denotes the macro-level loss of the $i^\text{th}$ phrase 
\begin{align}
L_{\text{mac}}^i := \begin{cases}
0 & i=1\\
T_{m'\rightarrow n'} & i=2,\cdots,p
\end{cases}
\end{align}
Here,  $T_{m'\rightarrow n'}$ is the transition loss defined the same way as in Section \ref{Meso-level-loss}. $m'$ and $n'$ index progression candidates to be connected. $p$ is the total number of phrases.% (assuming the chords of the very two bars connecting the two phrases are  $r_{m'}^{\text{chord}}$ and $r_{n'}^{\text{chord}}$) .

Finally, we define the total loss of the $s^{\text{th}}$ candidate of the $i^{\text{th}}$ phrase by a weighted sum
\begin{align}
\begin{split}
L^{i,s}_{\text{total}} = (\beta &(1 - L^{i,s}_{\text{mic}}) + (1-\beta) (1 - L^{i,s}_{\text{mes}})) \\
&+ \max_{t}{\{L^{i-1, t}_{\text{total}} + \alpha (1 - L^{i,s}_{\text{mac}})\}},
\end{split}
\end{align}
where $\alpha$, $\beta$ are parameters that we can tune between 0 and 1. Then we use DP to integrate the three levels of losses and search for the optimal chord progressions which minimize the total loss $L^{i}_{\text{total}}$  at $i = p$.

\subsection{An Overview of AccoMontage}
Based on the harmonization results from the last section, we apply AccoMontage \cite{zhao2021accomontage} to generate complete piano accompaniments in full length. The AccoMontage system offers a hybrid search-style transfer methodology for accompaniment arrangement: given a lead melody together with a chord progression inferred from Section \ref{harmonization_module}, it first searches for reference pieces of accompaniments by dynamic programming. It then re-harmonizes the reference accompaniment to the given chord progression by deep learning-based style transfer. Such a pipeline is inspired by common practice that delicate music textures can often be applied in bulk, instead of composing from scratch.

Specifically, reference accompaniment pieces are searched phrase by phrase based on: 1) phrase-level fitness to the melody, and 2) transition smoothness between consecutive phrases. The overall searching process is optimized by the Viterbi algorithm \cite{forney1973viterbi}. The re-harmonization is implemented with a VAE framework which is capable of chord-texture disentanglement \cite{wang2020learning}. By varying the chord representation, we can re-harmonize the accompaniment while keeping its texture. The whole pipeline of AccoMontage secures a delicate accompaniment arrangement with coherent and structured texture. We refer readers to the original work \cite{zhao2021accomontage} for more technical details. %In this workour case, we re-generate the reference accompaniment with the representation of our chord progression derived by the harmonization module in Section 3.2.

\subsection{Graphical User Interface for Controllability}

AccoMontage2 provides a GUI for users to select styles of harmonized chords and accompaniment textures. The process begins with uploading a MIDI file with a melody track and setting the original chord progression styles and piano texture styles. Three labels have to be assigned manually, including phrase boundaries, key, and mode (major or minor in the current version of the system). 

\begin{figure}[htb]
 \centerline{
 \includegraphics[width=0.99\columnwidth]{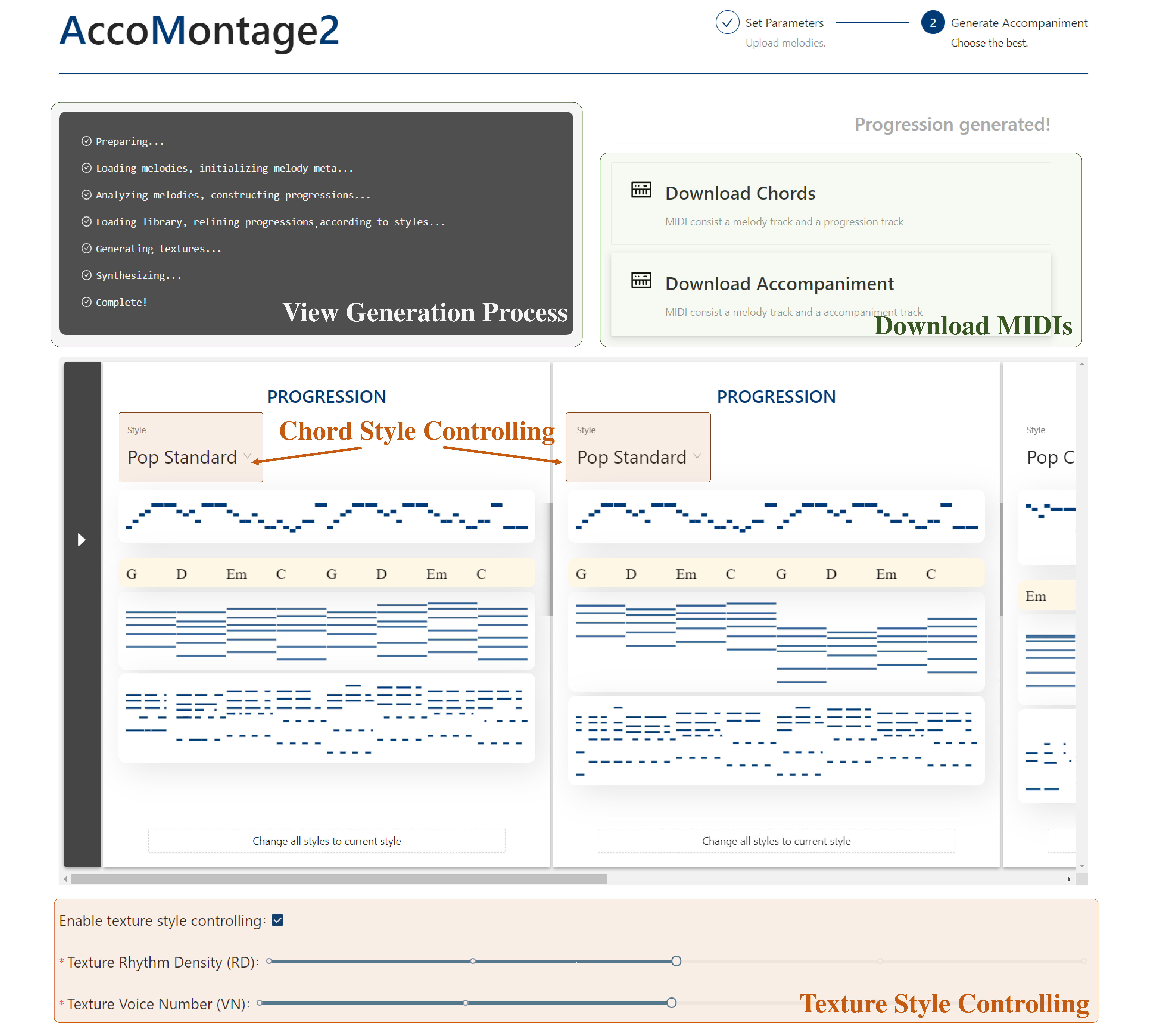}}
 \caption{A screenshot of the Graphical User Interface.}
 \label{gui_screenshot}
\end{figure}

The system will then proceed to generation.  When the generation is finished, users are able to: 1) listen to and download the generated audio; 2) select a new harmonization style for each individual phrase or the whole; 3) reset the texture style and re-generate the accompaniment. Figure \ref{gui_screenshot} shows the GUI interface of AccoMontage2.

\section{Generation Results}

%\begin{figure*}[htb]
%    \centering
%    \includegraphics[width=2\columnwidth]{figs/section4_SPACE_TEST.png}
%    \caption{SPACE PLACEHOLDER!!!}
%    \label{fig:my_label}
%\end{figure*}

\begin{figure*}[htb]
    \centering
    \includegraphics[width=1.95\columnwidth]{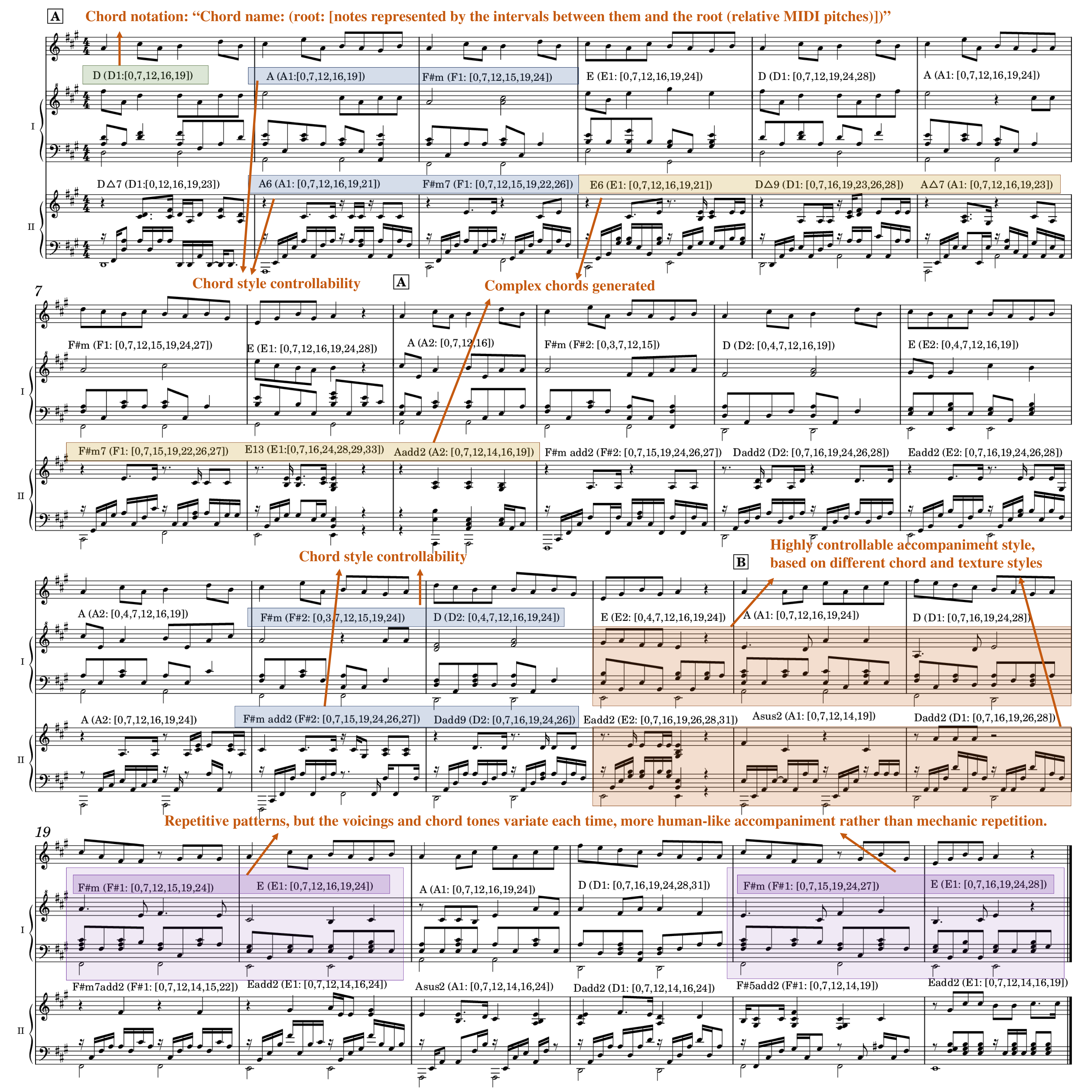}
    \caption{Harmonization and accompaniment arrangement results for \textit{Dinners 1} from the Nottingham Dataset. The 24-bar melody has an \texttt{A8A8B8} phrase structure, and the AccoMontage2 system achieves a high degree of style controllability.}
    \label{fig:sheet}
\end{figure*}

In this section, we showcase one long-term generation result of the AccoMontage2 system. We set $\alpha=0.1$ and $\beta=0.5$ in the harmonization algorithm and test our model on a 24-bar melody piece with phrase label \verb|A8A8B8|. The first track in Figure \ref{fig:sheet} shows the original melody piece and the other two tracks are two different versions of accompaniment generation.

The result shows that AccoMontage2 is able to achieve high controllability in chord style and texture style. On the top of track two (I) and track three (II) in Figure \ref{fig:sheet}, we present using chord notations the harmonization results of chord style ``Pop-standard'' and ``Pop-complex'' respectively. Track two and track three are the whole accompaniment results. While track two (I) is based on ``Pop-standard'' and a sparse texture style, track three (II) is based on ``Pop-complex'' and a dense texture style. We see that both results match with the melody in harmonicity and are able to provide chord and texture variations. 

\section{Evaluation}
We conduct two comparative experiments to validate our AccoMontage2 system, one for harmonization and the other for accompaniment arrangement. We first show the dataset and the baseline model in Section \ref{dataset} and \ref{baseline_method}. Then we present the experimental results in Section \ref{harmonization_results} and \ref{arrangement_results}. Audio examples of our proposed system and ablation studies are available via our GUI link.

\subsection{Dataset}\label{dataset}
For the harmonization experiment, we use Nottingham Dataset \cite{nottingham}, which provides around 1000 pairs of the query lead melody and the ground-truth chords. For arrangement generation experiments, we use the POP909 Dataset \cite{pop909-ismir2020}, in which each piece contains a lead melody, annotated chords, and a piano accompaniment. For both data sources, we first select the pieces of meter $\frac{2}{4}$ and $\frac{4}{4}$ and then randomly sample 6 pieces for our experiment. We manually annotate their phrase segmentation and only allow the phrase length of 4 bars and 8 bars.

\subsection{Baseline Method}\label{baseline_method}
We apply the chord generation model \cite{lim2017chord} based on bidirectional long short-term memory networks (BLSTM) \cite{DBLP:journals/neco/HochreiterS97} as our baseline model. This model takes a symbolic melody with the information of time signature, measure, and key as input, and outputs a harmonization result of a sequence of major and minor triads. The BLSTM model considers temporal dependencies by storing both past and future information, reflecting musical context in both forward and backward directions, It is trained on a lead sheet database provided by Wikifonia.org and has achieved reasonable results quantitatively and qualitatively.

\subsection{Harmonization Results}\label{harmonization_results}
% \subsubsection{Objective Evaluation}

% We collect the harmonization result of our model and the LSTM baseline, and also the original harmony pieces. We use the DP algorithm to calculate the multi-level loss on these pieces. Table 1 shows these sub-terms together with the total DP loss of the three methods respectively. Results show that our model outperforms the LSTM baseline and the original piece.

% \begin{table}[htbp]
%   \centering
%     \resizebox{.46\textwidth}{!}{
%     \begin{tabular}{lcccc}
%     \toprule
%           & \textbf{Micro-\textit{Lv.}} & \textbf{Meso-\textit{Lv.}} & \textbf{Macro-\textit{Lv.}} & \textbf{Total} \\
%     \midrule
%     \textbf{Original}   & \textbf{0.19} & 0.50 & 0.56 & 0.37 \\
%     \textbf{LSTM}       & 0.27 & 0.21 & 0.23 & 0.24 \\
%     \textbf{Ours}       & 0.29 & \textbf{0.16} & \textbf{0.22} & \textbf{0.23} \\
%     \bottomrule
%     \end{tabular}%
%     }
%   \caption{Comparison of multi-level harmonization score.}
%   \label{multi_level_score}%
% \end{table}%

% \subsubsection{Subjective Evaluation}\label{subjective_harmonization_section}
We conduct a survey to evaluate the harmonization performance of our model. Our survey has 6 groups of harmonization results and each subject is required to listen to 3 (chosen randomly). Within each group, the subject first listens to a single melody. Each melody is a full-length song randomly selected from the Nottingham Dataset, with an average length of 32 bars (64 seconds). We harmonize the melody using our model and the BLSTM baseline, and additionally acquire an original harmonization using the ground truth chord labels. The subjects are then required to evaluate all three versions of harmonization. The rating is based on a five-point scale from 1 (very poor) to 5 (very high) according to three criteria:
\begin{enumerate}
    \item \textbf{Harmonicity}: How well do the melody and chords stay in harmony with each other;  
    \item \textbf{Creativity}: How creative the harmonization is;
    \item \textbf{Musicality}: The overall musicality.
\end{enumerate}

\begin{figure}[htb]
 \centerline{
 \includegraphics[width=0.95\columnwidth]{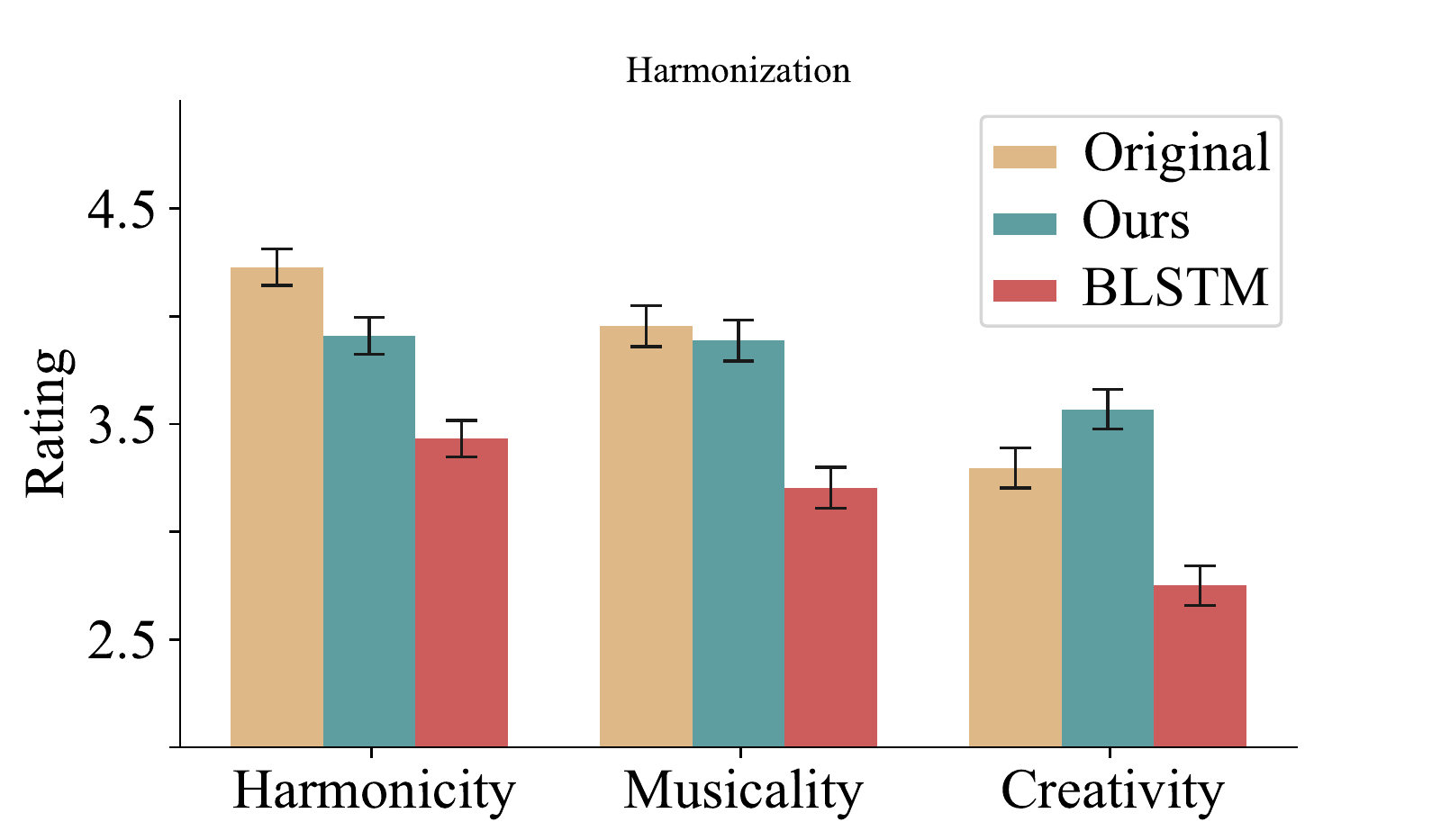}}
 \caption{Subjective evaluation for melody harmonization.}
 \label{Harmonization_Subject_Evaluation}
\end{figure}

A total of 15 subjects with diverse musical backgrounds participated in our survey and we obtain 44 effective ratings for each criterion. As shown in Figure \ref{Harmonization_Subject_Evaluation}, the height of the bars denotes the mean values of the ratings. The error bars stand for the mean square errors (MSEs) computed via within-subject ANOVA\cite{scheffe1999analysis}. For harmonicity, the original receives the best rating, while our model performs significantly better than the BLSTM baseline. As for the overall musicality, our model is comparable with the original harmonization. For creativity, our model reaches the best. Note that our model supports complex harmonization with voice leading and ninth chords, while the original and the baseline support up to seventh chords in plain root position. Such evaluation results demonstrate that our model introduces \textit{tension} to ``flavor'' the music while the overall musicality is not affected. For all three criteria, the rating results are statistically significant (p-value $p < 0.05$).

\subsection{Accompaniment Generation Results}\label{arrangement_results}
We conduct another survey to evaluate our model in terms of overall accompaniment generation. In our survey, each subject still listens to 3 groups of generation results (randomly chosen from 6 groups). Within each group, the subjects first listen to a 32-bar melody randomly selected from the POP909 Dataset. Our model generates piano accompaniment through the complete AccoMontage2 pipeline. For the BLSTM baseline, we feed its harmonization results to AccoMontage and obtain the baseline accompaniment. As POP909 contains piano arrangements created by professional musicians for each song, we also have the original accompaniment. The subjects are then required to evaluate all three versions of accompaniment based on the same scale and criteria as in Section \ref{harmonization_results}.

\begin{figure}[htb]
 \centerline{
 \includegraphics[width=0.95\columnwidth]{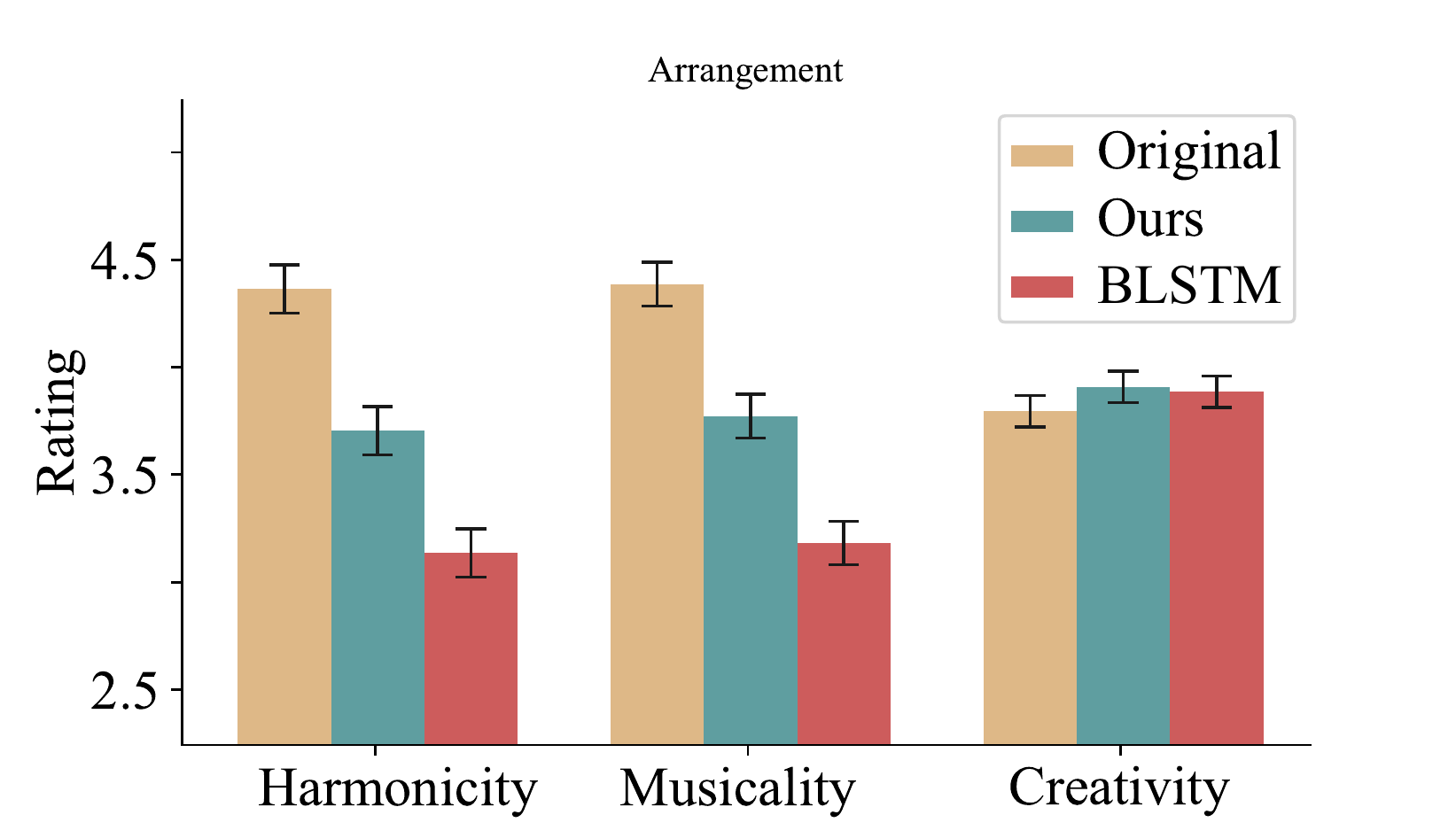}}
 \caption{Subjective evaluation for accompaniment arrangement.}
 \label{Arrangement_Subject_Evaluation}
\end{figure}

We collect a total of 44 effective ratings for each criterion. Figure \ref{Arrangement_Subject_Evaluation} shows the evaluation results in the same format as in Section \ref{harmonization_results}. We report a significantly better performance of our model compared with the BLSTM baseline in harmonicity and musicality ($p<0.05$), and a marginally better performance than both the baseline and the original in creativity.

\section{Conclusion and Future Work}

In conclusion, we contribute a pipeline of algorithms to automatically harmonize and arrange piano accompaniments for melodies of whole-piece popular and folk songs. The system is named after AccoMontage2, built upon its original version which uses a hybrid approach to select accompaniment candidates, edit the candidates using style transfer, and then concatenate them into an organic whole. AccoMontage2 contains two novel modules: a state-of-the-art harmonizer and a GUI for controllable arrangement via interfering in the harmonization and arrangement styles.

In the future, we plan to further optimize the arrangement pipeline by: 1) automatically labeling the melody phrase, 2) extending the model capability to deal with triple meters, and 3) exploring full-band arrangement.

%\newpage
% For bibtex users:
\bibliography{ISMIRtemplate}

\begin{thebibliography}{10}
\providecommand{\url}[1]{#1}
\csname url@samestyle\endcsname
\providecommand{\newblock}{\relax}
\providecommand{\bibinfo}[2]{#2}
\providecommand{\BIBentrySTDinterwordspacing}{\spaceskip=0pt\relax}
\providecommand{\BIBentryALTinterwordstretchfactor}{4}
\providecommand{\BIBentryALTinterwordspacing}{\spaceskip=\fontdimen2\font plus
\BIBentryALTinterwordstretchfactor\fontdimen3\font minus
  \fontdimen4\font\relax}
\providecommand{\BIBforeignlanguage}[2]{{%
\expandafter\ifx\csname l@#1\endcsname\relax
\typeout{** WARNING: IEEEtran.bst: No hyphenation pattern has been}%
\typeout{** loaded for the language `#1'. Using the pattern for}%
\typeout{** the default language instead.}%
\else
\language=\csname l@#1\endcsname
\fi
#2}}
\providecommand{\BIBdecl}{\relax}
\BIBdecl

\bibitem{wang2020learning}
Z.~Wang, D.~Wang, Y.~Zhang, and G.~Xia, ``Learning interpretable representation
  for controllable polyphonic music generation,'' in \emph{Proceedings of the
  21th International Society for Music Information Retrieval Conference,
  {ISMIR}}, 2020, pp. 662--669.

\bibitem{xia_2018}
\BIBentryALTinterwordspacing
G.~Xia, ``Expressive collaborative music performance via machine learning,''
  Ph.D. dissertation, Carnegie Mellon University, {USA}, 2016. [Online].
  Available: \url{https://doi.org/10.1184/r1/6716609.v1}
\BIBentrySTDinterwordspacing

\bibitem{DBLP:conf/ismir/LinXKJ21}
L.~Lin, G.~Xia, Q.~Kong, and J.~Jiang, ``A unified model for zero-shot music
  source separation, transcription and synthesis,'' in \emph{Proceedings of the
  22nd International Society for Music Information Retrieval Conference,
  {ISMIR}}, 2021, pp. 381--388.

\bibitem{zhao2021accomontage}
J.~Zhao and G.~Xia, ``Accomontage: Accompaniment arrangement via phrase
  selection and style transfer,'' in \emph{Proceedings of the 22nd
  International Society for Music Information Retrieval Conference, {ISMIR}},
  2021, pp. 833--840.

\bibitem{chuan2007hybrid}
C.-H. Chuan, E.~Chew \emph{et~al.}, ``A hybrid system for automatic generation
  of style-specific accompaniment,'' in \emph{Proceedings of the 4th
  international joint workshop on computational creativity}.\hskip 1em plus
  0.5em minus 0.4em\relax Goldsmiths, University of London London, 2007, pp.
  57--64.

\bibitem{simon2008mysong}
I.~Simon, D.~Morris, and S.~Basu, ``Mysong: automatic accompaniment generation
  for vocal melodies,'' in \emph{Proceedings of the SIGCHI conference on human
  factors in computing systems}, 2008, pp. 725--734.

\bibitem{tsushima2017function}
H.~Tsushima, E.~Nakamura, K.~Itoyama, and K.~Yoshii, ``Function- and
  rhythm-aware melody harmonization based on tree-structured parsing and
  split-merge sampling of chord sequences,'' in \emph{Proceedings of the 18th
  International Society for Music Information Retrieval Conference, {ISMIR}},
  2017, pp. 502--508.

\bibitem{tsushima2018generative}
------, ``Generative statistical models with self-emergent grammar of chord
  sequences,'' \emph{Journal of New Music Research}, vol.~47, no.~3, pp.
  226--248, 2018.

\bibitem{lim2017chord}
H.~Lim, S.~Rhyu, and K.~Lee, ``Chord generation from symbolic melody using
  {BLSTM} networks,'' in \emph{Proceedings of the 18th International Society
  for Music Information Retrieval Conference, {ISMIR}}, 2017, pp. 621--627.

\bibitem{yeh2021automatic}
Y.-C. Yeh, W.-Y. Hsiao, S.~Fukayama, T.~Kitahara, B.~Genchel, H.-M. Liu, H.-W.
  Dong, Y.~Chen, T.~Leong, and Y.-H. Yang, ``Automatic melody harmonization
  with triad chords: A comparative study,'' \emph{Journal of New Music
  Research}, vol.~50, no.~1, pp. 37--51, 2021.

\bibitem{sun2021melody}
C.-E. Sun, Y.-W. Chen, H.-S. Lee, Y.-H. Chen, and H.-M. Wang, ``Melody
  harmonization using orderless nade, chord balancing, and blocked gibbs
  sampling,'' in \emph{ICASSP 2021-2021 IEEE International Conference on
  Acoustics, Speech and Signal Processing (ICASSP)}.\hskip 1em plus 0.5em minus
  0.4em\relax IEEE, 2021, pp. 4145--4149.

\bibitem{bretan2016unit}
M.~Bretan, G.~Weinberg, and L.~Heck, ``A unit selection methodology for music
  generation using deep neural networks,'' \emph{arXiv preprint
  arXiv:1612.03789}, 2016.

\bibitem{chen2013automatic}
P.-C. Chen, K.-S. Lin, and H.~H. Chen, ``Automatic accompaniment generation to
  evoke specific emotion,'' in \emph{2013 IEEE International Conference on
  Multimedia and Expo (ICME)}.\hskip 1em plus 0.5em minus 0.4em\relax IEEE,
  2013, pp. 1--6.

\bibitem{wu2016emotion}
Y.-C. Wu and H.~H. Chen, ``Emotion-flow guided music accompaniment
  generation,'' in \emph{2016 IEEE International Conference on Acoustics,
  Speech and Signal Processing (ICASSP)}.\hskip 1em plus 0.5em minus
  0.4em\relax IEEE, 2016, pp. 574--578.

\bibitem{liu2012polyphonic}
C.-H. Liu and C.-K. Ting, ``Polyphonic accompaniment using genetic algorithm
  with music theory,'' in \emph{2012 IEEE Congress on Evolutionary
  Computation}.\hskip 1em plus 0.5em minus 0.4em\relax IEEE, 2012, pp. 1--7.

\bibitem{fujishima1999real}
T.~Fujishima, ``Realtime chord recognition of musical sound: a system using
  common lisp music,'' in \emph{Proceedings of the 1999 International Computer
  Music Conference, {ICMC}}.\hskip 1em plus 0.5em minus 0.4em\relax Michigan
  Publishing, 1999.

\bibitem{dong2018musegan}
H.-W. Dong, W.-Y. Hsiao, L.-C. Yang, and Y.-H. Yang, ``Musegan: Multi-track
  sequential generative adversarial networks for symbolic music generation and
  accompaniment,'' in \emph{Proceedings of the AAAI Conference on Artificial
  Intelligence}, vol.~32, no.~1, 2018.

\bibitem{liu2018lead}
H.-M. Liu and Y.-H. Yang, ``Lead sheet generation and arrangement by
  conditional generative adversarial network,'' in \emph{2018 17th IEEE
  International Conference on Machine Learning and Applications (ICMLA)}.\hskip
  1em plus 0.5em minus 0.4em\relax IEEE, 2018, pp. 722--727.

\bibitem{jia2019impromptu}
B.~Jia, J.~Lv, Y.~Pu, and X.~Yang, ``Impromptu accompaniment of pop music using
  coupled latent variable model with binary regularizer,'' in \emph{2019
  International Joint Conference on Neural Networks (IJCNN)}.\hskip 1em plus
  0.5em minus 0.4em\relax IEEE, 2019, pp. 1--6.

\bibitem{ren2020popmag}
Y.~Ren, J.~He, X.~Tan, T.~Qin, Z.~Zhao, and T.-Y. Liu, ``Popmag: Pop music
  accompaniment generation,'' in \emph{Proceedings of the 28th ACM
  International Conference on Multimedia}, 2020, pp. 1198--1206.

\bibitem{zhu2018xiaoice}
H.~Zhu, Q.~Liu, N.~J. Yuan, C.~Qin, J.~Li, K.~Zhang, G.~Zhou, F.~Wei, Y.~Xu,
  and E.~Chen, ``Xiaoice band: A melody and arrangement generation framework
  for pop music,'' in \emph{Proceedings of the 24th ACM SIGKDD International
  Conference on Knowledge Discovery \& Data Mining}, 2018, pp. 2837--2846.

\bibitem{kotoulas}
\BIBentryALTinterwordspacing
N.~Kotoulas, ``Supersize your midi collection.'' [Online]. Available:
  \url{https://www.pianoforproducers.com/nikos-ultimate-midi-pack/}
\BIBentrySTDinterwordspacing

\bibitem{minorloss}
L.~Trulla, N.~Di~Stefano, and A.~Giuliani, ``Computational approach to musical
  consonance and dissonance,'' \emph{Frontiers in Psychology}, vol.~9, p. 381,
  04 2018.

\bibitem{forney1973viterbi}
G.~D. Forney, ``The viterbi algorithm,'' \emph{Proceedings of the IEEE},
  vol.~61, no.~3, pp. 268--278, 1973.

\bibitem{nottingham}
E.~Foxley, ``Nottingham database,'' [EB/OL],
  \url{https://ifdo.ca/~seymour/nottingham/nottingham.html} Accessed May 25,
  2021.

\bibitem{pop909-ismir2020}
Z.~Wang*, K.~Chen*, J.~Jiang, Y.~Zhang, M.~Xu, S.~Dai, G.~Bin, and G.~Xia,
  ``Pop909: A pop-song dataset for music arrangement generation,'' in
  \emph{Proceedings of 21st International Conference on Music Information
  Retrieval, {ISMIR}}, 2020.

\bibitem{DBLP:journals/neco/HochreiterS97}
S.~Hochreiter and J.~Schmidhuber, ``Long short-term memory,'' \emph{Neural
  Computation}, vol.~9, no.~8, pp. 1735--1780, 1997.

\bibitem{scheffe1999analysis}
H.~Scheffe, \emph{The analysis of variance}.\hskip 1em plus 0.5em minus
  0.4em\relax John Wiley \& Sons, 1999, vol.~72.

\end{thebibliography}

% For non bibtex users:
%\begin{thebibliography}{citations}
% \bibitem{Author:17}
% E.~Author and B.~Authour, ``The title of the conference paper,'' in {\em Proc.
% of the Int. Society for Music Information Retrieval Conf.}, (Suzhou, China),
% pp.~111--117, 2017.
%
% \bibitem{Someone:10}
% A.~Someone, B.~Someone, and C.~Someone, ``The title of the journal paper,''
%  {\em Journal of New Music Research}, vol.~A, pp.~111--222, September 2010.
%
% \bibitem{Person:20}
% O.~Person, {\em Title of the Book}.
% \newblock Montr\'{e}al, Canada: McGill-Queen's University Press, 2021.
%
% \bibitem{Person:09}
% F.~Person and S.~Person, ``Title of a chapter this book,'' in {\em A Book
% Containing Delightful Chapters} (A.~G. Editor, ed.), pp.~58--102, Tokyo,
% Japan: The Publisher, 2009.
%
%
%\end{thebibliography}

\end{document}